\documentclass[12pt]{article}
\usepackage[T1]{fontenc}
\usepackage[a4paper,top=3.5cm, bottom=5.5cm,left=2.5cm,right=2.5cm]{geometry}
\usepackage{graphicx}
\usepackage[utf8]{inputenc}
\usepackage{amsmath}
\usepackage{slashed}
\usepackage{mathtools}
\usepackage{amsthm}
\usepackage{amsfonts}
\usepackage{amssymb}% http://ctan.org/pkg/amssymb
\usepackage{pifont}% http://ctan.org/pkg/pifont
\newcommand{\cmark}{\ding{51}}%
\newcommand{\xmark}{\ding{55}}%
\usepackage{booktabs}
\usepackage[english]{babel}
\usepackage{setspace}
\usepackage{geometry}
\usepackage{xcolor}
\begin{document}
\thispagestyle{empty}
$\,$

\vspace{32pt}
\begin{center}
% \maketitle
%\textbf{\Large  Radiative corrections for $S \to ZZ$ and $S \to W^+W^-(\gamma)$ in the real singlet extension of the standard model} 
\textbf{\Large  Radiative corrections of heavy scalar decays to gauge bosons in the singlet extension of the Standard Model} 

\vspace{30pt}
G.~Ria$^a$, D. Meloni$^a$
\vspace{16pt}

\textit{$^a$Dipartimento di Matematica e Fisica, 
Universit\`a di Roma Tre}\\
\textit{INFN, Sezione di Roma Tre, 00146 Rome, Italy}\\
\vspace{16pt}

\texttt{gabrieleria.uniroma3@gmail.com, meloni@fis.uniroma3.it}
\end{center} 
 \abstract
Assuming the existence of a new real scalar singlet $s^0$ coupled to the Standard Model via a scalar quartic portal interaction, we 
compute the radiative corrections to the decay rates of the heavy scalar mass eigenstate to a couple of gauge bosons $ZZ$ and $W^+W^-(\gamma)$, showing that they 
can give a contribution as large as ${\cal{O}}$(5\%) and ${\cal{O}}$(7\%), respectively.
% for heavy scalar masses around $m_S \sim 250$ GeV, respectively.   
We also explicitly analyze in detail their dependence on the heavy mass $m_S$ and on the scalar mixing angle $\alpha$, finding that, especially in the large-mass region, these depend on the sign and the assumed value of $\sin\alpha$.
\section{Introduction}
In June of 2012, the LHC experiment \cite{atlas1,cms1} has finally completed the spectrum of the Standard Model with the discovery of the Higgs 
boson, predicted in the 60's by Higgs \cite{Higgs,Higgs1}, Englert, Brout \cite{Brout}, Guralnik, Hagen and Kibble \cite{GHK}. However, 
the structure and the physics behind the Higgs sector are not completely clear and this represents a possible gateway to 
the manifold conceivable extensions of the Standard Model (SM). 
One of the simplest renormalizable enlargement  of the Higgs sector is 
constructed by adding to the SM Lagrangian one additional spinless real singlet under all groups of the SM, which develops its own vacuum expectation
value \cite{ssm1,ssm3,ssm4,status,ultimoRobens,Lewis1,Lewis2,Lewis3}.\\
Beside being easy to implement, the physics of a scalar singlet has received a lot of attention in the recent years for several reasons; among them, it can help in solving the issues related to the  metastability  of  the  electroweak
vacuum  \cite{Buttazzo:2013uya, strumia} if the Higgs potential receives a correction due to new physics
which modifies it at large  field values \cite{Groos} and it could provide a door to hidden sectors \cite{ssm2} to which it is coupled.
The singlet model has the advantage of depending on relatively few parameters and this implies a feasible experimental study at the LHC for the analysis of the new physic effects in the Higgs boson couplings, 
searches for heavy SM-like Higgs bosons \cite{ATLAS:2014aga,Khachatryan:2015cwa} 
and direct searches for resonant di-Higgs production \cite{Aad:2015uka,Aad:2014yja,Khachatryan:2015yea}; in the absence of linear and triple self-interactions, this model possesses a $\mathbb{Z}_2$-symmetry and 
the singlet can be a viable candidate for dark matter, although for masses somehow larger than 500 GeV \cite{Barger:2007im, Casas:2017jjg}
the couplings of the dark matter to the known particles occur only through the mixing of the singlet field with the SM Higgs boson.
Without a $\mathbb{Z}_2$-symmetry a strong first order electroweak phase transition is allowed and additional sources of ${\cal{CP}}$ violation occur in the scalar potential.
In this article we limit ourselves to a situation where the new singlet $s^0$ communicates with the $SU(2)_L$ doublet $\phi$ only via a quartic interaction of the form, $$\kappa\,(\phi^\dagger\phi) (s^0)^2\,.$$
This implies that the would-be Higgs boson of the SM mixes with the new singlet leading to the existence of two mass eigenstates, the lighter of which ($H$) is 
the experimentally observed Higgs boson whereas the heaviest one ($S$) is a new state not 
seen so far in any collider experiments. We call this model the  {\it Singlet Extension of the SM} ({\bf SSM}).
Since only $\phi$ is coupled to ordinary matter, the main production mechanisms and decay channels of $H$ and $S$ are essentially the same as those
of the usual SM Higgs particle, with couplings rescaled by quantities which depend on the scalar mixing angle, called $\alpha$, whose bounds 
have been discussed in details in \cite{status,ultimoRobens,deltaR,Run2}. We will focus our attention on the mass range: $200 \leq m_S \leq 1000$ GeV.
For masses larger than $\gtrsim 200$ GeV, the most important decay channels of the heavy state $S$ are those 
to a pair of vector bosons $S \to VV$ and, when kinematically allowed ($m_S > 2\,m_H$), to a pair of lighter scalars and top quarks, $S \to HH, \bar{t}t$.
With the run II at LHC, the exploration of the scalar sector is expected to reveal more details.
Thus the comparison between theory and data requires precise predictions obtained through higher-order calculations.
To this aim, we evaluated the radiative corrections to the decay rates and studied in details their dependence on the singlet mass $m_S$ as well as on the mixing angle $\alpha$.
For the non-diagonal scalar sector, we use the gauge-independent renormalization scheme called "\textit{improved on-shell}" (iOS) \cite{ultimo2Robens}
while for the remaining %electroweak 
quantities we work within the standard "\textit{on-shell}" renormalization scheme (OS).\\
% We have analysed the impact of the gauge dependence on the decay rates and found that they are less than  $\sim |3|\%$ in both channels.\\
The main result of this paper is that below $m_S \lesssim 700$ GeV, the above mentioned corrections are positive and reach a maximum of ${\cal{O}}$(5\%) in the $ZZ$ channel and ${\cal{O}}$(7\%) in the $ W^+W^-(\gamma)$ channel (both for $m_S \sim 200$ GeV),
almost independently on the mixing angle $\alpha$, whereas for larger masses the one-loop contributions drive the decay rates to smaller values with a more pronounced 
dependence on $\alpha$. \\
The structure of the paper is as follows: in Sect.\ref{desmod} we remind the reader of the relevant features of the Singlet Standard Model;
in Sect.\ref{SSMrenorm} we illustrate the details of our renormalization procedure that we apply 
in Sect.\ref{SVVvertex} to discuss the structure of the $SVV$ renormalized vertices.
The radiative corrections to the $S\to VV $ decay rates are illustrated in detail in Sect.\ref{SVVdecay}; 
Sect.\ref{concl} is devoted to our conclusions.

\section{Description of the Singlet Standard Model}
\label{desmod}
The scalar potential of the model analyzed here is given by the usual SM potential $V_{\textup{sm}}(\phi)$, with $\phi$ representing the 
SM scalar field, augmented with the new contributions due to quadratic and quartic terms of the new singlet $s^{0}$,
and a portal interaction among $s^{0}$ and $\phi$, contained in 
$V_{\textup{np}}(\phi,s^0)$, as specified below:
\begin{align}
\nonumber
V_{\textup{sm}}(\phi) \ =& \ \mu^2 (\phi^\dagger\phi) + \lambda (\phi^\dagger\phi)^2, \\ 
V_\textup{np}(\phi,s^0) \ =& \ \mu_{s}^2\,(s^0)^2 + \rho\,(s^0)^4 + \kappa\,(\phi^\dagger\phi) (s^0)^2\,. \nonumber
\end{align}
The potential $V_\textup{np}(\phi,s^0)$
exhibits a $\mathbb{Z}_2$-symmetry under which $s^0 \rightarrow -s^0$, so 
that linear and trilinear terms in $s^0$ are not allowed \footnote{ A similar treatment to the one discussed in this work, including the linear and trilinear terms, can be found in \cite{ kanemura1, kanemura2, kanemura3}.}.
Since $s^{0}$ is a true isospin singlet, the masses of the W and Z gauge bosons  
are only due to the coupling with the $\phi$ fields that we parametrize as
$\phi=[\eta^+,\phi^0+i \eta_3/\sqrt{2}]^{T}$,  where $\phi^0= (v+h) /\sqrt{2}$ ($v=246.22$ GeV) and $\eta^\pm=(\eta_1 \pm i \eta_2)/\sqrt{2}$ and $\eta_3$ are the Goldstone bosons.
Beside the vev of the $\phi$ field, we also consider the possibility that $s^0$ acquires a non-zero vacuum expectation value (vev) $w$,
and thus the expansion of the field around its classical minimum is set as $s^0= (w +s)/\sqrt{2}$.
The most immediate consequence of the potential  $\mathcal{V}_{\textup{sm}}(\phi)+\mathcal{V}_\textup{np}(\phi,s^0)$ is that a non-diagonal mass matrix is generated for 
the two neutral states $h$ and $s$,
\begin{equation}
\label{matrmassa}
\mathcal{M}^2=\begin{pmatrix}
2 \lambda v^2 & \kappa v w \\ 
\kappa v w & 2 \rho w^2
\end{pmatrix}.
\end{equation}
Considering all couplings as real parameters, the positivity of the mass matrix is ensured requiring that \cite{Groos}
\begin{equation}
\lambda>\frac{\kappa^2}{4\rho} \ \ , \ \ \rho>0.
\end{equation}
The symmetric mass matrix in eq.(\ref{matrmassa}) is diagonalized by an orthogonal transformation which, in turn, realizes a 
mapping between the  Lagrangian states and the physical fields $H$ and $S$ \cite{Groos,ultimo2Robens} :
% \begin{equation}
\begin{align}
m_{H,S}^2 \ = \ &\lambda v^2 + \rho w^2 \mp \frac{\rho w^2-\lambda v^2}{\cos 2\alpha}, \\
\phi^0 \ = \  &\frac{1}{\sqrt{2}}(v + H\cos\alpha + S\sin\alpha), \\
s^0 \ = \ &\frac{1}{\sqrt{2}}(w - H\sin\alpha + S\cos\alpha),
\end{align}
% \end{equation}
where the mixing angle $\alpha \in [-\pi/2,\pi/2]$. In the rest of this paper we consider the  $H$ field as the lightest 
mass eigenstate and we identify it with the SM Higgs boson, so we always consider $sign\left(\rho w^2-\lambda v^2\right)\times 
sign\left(\cos 2\alpha \right) >0$.
Expanding $\mathcal{V}_{\textup{np}}(\phi,s^0)$ in terms of the physical fields, we get the scalar trilinear and quartic couplings which are reported in App.\ref{SCALCOU}.
The mixing angle $\alpha$ can be expressed in terms of the model parameters and vevs so that
\begin{equation}
\label{angletan}
\tan2\alpha=\frac{\kappa v w}{\rho w^2 - \lambda v^2}\,.
\end{equation}
Here we limit ourselves to the  mass range  $200 \leq m_S \leq 1000$ GeV, which corresponds to 
the bound $0.018 \leq |\sin\alpha| \leq 0.36$  \cite{ultimo2Robens}. 
Notice that in the decoupling limit $(v/w) \ll 1$, the expressions for the masses and mixing are well approximated by:
\begin{equation}
\label{limitdef}
m_H^2 \equiv 2v^2 \lambda_\textrm{SM} \simeq \ 2v^2\left(\lambda - \frac{\kappa^2}{4\rho}\right) \ \ \ , \ \ \ m_S^2 \simeq \ 2\rho w^2 + \frac{\kappa^2 v^2}{2\rho} \ \ \ , \ \ \ \sin\alpha \simeq \ \frac{\kappa v }{2 \rho w}\,,
\end{equation}
which clearly shows that the quartic coupling $\lambda_\textrm{SM}$ receives a correction proportional to the ratio among the portal coupling $\kappa$ 
and the quartic of the $s^0$ field \cite{strumia}.
The couplings of the $H$ and $S$ fields with gauge bosons and fermions are similar to the ones of the SM Higgs, rescaled by the appropriate mixing \cite{Groos}:
\begin{align}
\mathcal{L}_{\textrm{SSM}} \ &\supset \ \frac{H c_\alpha + S s_\alpha}{v}\left[m_Z^2 Z_\mu Z^\mu + 2m_W^2 W_\mu^+W^{\mu-} - \sum_f m_f \bar{f}f \right],
\end{align}
with $s_\alpha = \sin\alpha$ and $c_\alpha = \cos\alpha$.
Then the tree-level amplitude for the $S\to VV$ decays is given by:
\begin{eqnarray}
\label{couszz}
S(k)V(p,a)V(q,b) \ &\Longrightarrow& \ \rho_V \times [g^{\mu\nu}\epsilon_\mu^a(p)\epsilon_\nu^b(q)] \,,
% \end{align}
\end{eqnarray}
where $k,\,p$ and $q$ are the four momenta %of the vector bosons
and $a,b$ the polarizations. Here $\rho_V$ is the SSM bare coupling defined as
\begin{equation}
\label{rho}
\rho_V= e \frac{m_V^2}{s_W m_W} s_\alpha\,,
\end{equation}
where $V= W^{\pm} , Z$ and  $s_W=\sin \theta_W$, $\theta_W$ being the Weinberg's angle.
\section{Generalities on the renormalization procedure}
\label{SSMrenorm}
We start introducing the main renormalized quantities and counterterms of our interest \cite{Denner}:
\begin{align}
{\cal V}_0 \ =& \ {\cal V}(1 + \delta {\cal V})\,, \\
(m_V^2)_0 \ =& \ m_V^2 + \delta m_V^2\,, \\
(\theta_W)_0 \ =& \ \theta_W + \delta \theta_W\,, \\
\alpha_0 \ =& \ \alpha + \delta \alpha\,, \\
e_0 \ =& \ (1 + \delta Z_e) e\,,
\end{align}
where ${\cal V}_0$ is a short-hand notation for a generic coupling and $e$ is the electric charge.
The relevant difference with respect to the SM renormalization procedure is the presence of the mixing in the scalar sector. 
Splitting the bare mixing angle as $\alpha_0 \to \alpha + \delta \alpha$, the two physical scalar fields $S$ and $H$ are 
related to the bare ones via mixing specified as
\begin{align}
\label{mixmat}
\Bigg(\begin{matrix}
H_0\\
S_0
\end{matrix}\Bigg) &\to \begin{pmatrix}
1+\frac{\delta Z_H}{2} & \frac{\delta Z_{HS}}{2} - \delta \alpha \\ 
\frac{\delta Z_{SH}}{2} + \delta \alpha & 1+\frac{\delta Z_S}{2}
\end{pmatrix}\Bigg(\begin{matrix}
H\\
S
\end{matrix}\Bigg)\,.
\end{align}
%where $\delta \alpha$ is the mixing angle counterterm which produces the loop-corrected mass matrix.
We will also need the field renormalization constants for $W^\pm$, $Z^0$ and $\gamma$ defined as
\begin{align}
& \ \ \ \ W_0^\pm = \left(1 + \frac{1}{2}\delta Z_W \right) W^\pm, \\
\Bigg(\begin{matrix}
Z_0\\
\gamma_0
\end{matrix}\Bigg) & = \begin{pmatrix}
 1+\frac{\delta Z_{Z}}{2} & \frac{\delta Z_{Z \gamma}}{2} - \delta \theta_W \\
\frac{\delta Z_{\gamma Z}}{2} + \delta \theta_W & 1+\frac{\delta Z_{\gamma}}{2}\\ 
\end{pmatrix}\Bigg(\begin{matrix}
Z \\
\gamma
\end{matrix}\Bigg)\,,
\end{align}
where in the last line we explicitly show the counterterms entering in the mixing matrix of the neutral gauge bosons.
Notice that we can rewrite $\delta \theta_W$ as $\delta s_W^2/(2s_Wc_W)$ using: $\delta \theta_W =\delta s_W/ c_W $ and $\delta s_W=\delta s_W^2/(2 s_W)$.
As usual, the counterterms are fixed by renormalization conditions \cite{Denner} (the "hat" will be used to indicate renormalized quantities).
%Obviously, 
The tadpole of the scalar fields \footnote{The tadpoles are given by the following relations: $T_h = \mu v^2 + v^3 \lambda + v w^2 \kappa /2 \,, \  T_s = \mu_s w^2 + w^3 \rho + v^2 w \kappa /2 \,. $} must also be shifted:
\begin{align}
\hat{T}_H \to T_H +\delta t_H \ \ \ , \ \ \ \hat{T}_S \to T_S +\delta t_S\,,
\end{align}
where $T_H$ and $T_S$ are related with the tadpoles in the gauge basis $T_h$ and $T_s$ by the mixing,
\begin{align}
\Bigg(\begin{matrix}
T_h\\
T_s
\end{matrix}\Bigg) &\to \begin{pmatrix}
c_\alpha& s_\alpha \\ 
-s_\alpha & \ c_\alpha
\end{pmatrix}\Bigg(\begin{matrix}
T_H\\
T_S
\end{matrix}\Bigg).
\end{align}
If we impose the renormalization conditions $\hat{T}_H = T_H + \delta t_H=0$ and $\hat{T}_S = T_S + \delta t_S=0$, 
no scalar one-point insertions need to be explicitly computed  since it is 
equivalent to require that $v$ and $w$ are the physical vacuum expectation values of the doublet and the singlet fields, respectively. 
The next conditions involve the renormalized one-particle irreducible two-point functions of the scalar and vector fields; 
in the 't Hooft-Feynman gauge (that will be used throughout the rest of this paper) we have:
\begin{align}
\label{self}
\hat{\Gamma}^{ii'}(k) =& \ i(k^2 - m_{i}^2)\delta^{ii'} + i\hat{\Sigma}^{ii'}(k^2)\,, \\
\hat{\Gamma}^{jj'}_{\mu\nu}(k) =& -ig^{\mu\nu}(k^2-m_{j}^2)\delta^{jj'}-i\left(g^{\mu\nu} - \frac{k^\mu k^\nu}{k^2} \right)\hat{\Sigma}^{jj'}_T(k^2) - 
i\frac{k^\mu k^\nu}{k^2}\hat{\Sigma}^{jj'}_L(k^2)\,,
\end{align}
where $m_{i/j}$ is the mass of the incoming particle and ($ii'$, $ jj'$) can be one of the combinations  $\{HH, SS, HS, SH\}$ and $\{ WW, ZZ, \gamma\gamma, \gamma Z, Z\gamma \}$, respectively.
$\hat{\Sigma}^{jj'}_T$ and $\hat{\Sigma}^{jj'}_L$ are the transverse and longitudinal contributions to the self-energies. Following Ref.\cite{Denner}, 
we can impose the following conditions on the self-energy functions in the OS 
renormalization scheme, in which all renormalization conditions are formulated
for the external fields on their mass shell \footnote{The renormalization conditions for the mixing in the scalar sector are discussed below eq.(\ref{count5}).}:
\begin{eqnarray}
\nonumber
Re \hat{\Sigma}^{HH}(m_H^2) &=&   0\,, \quad  Re \hat{\Sigma}^{'HH} (k^2) |_{k^2=m_H^2} =  0 \ , \\\nonumber
Re \hat{\Sigma}^{SS}(m_H^2) &=&   0\,, \quad  Re \hat{\Sigma}^{'SS} (k^2) |_{k^2=m_H^2} =  0 \ , \\\nonumber
\widetilde{Re} \hat{\Sigma}^{WW}_T(m_W^2) &=& 0\,, \quad \widetilde{Re} \hat{\Sigma}^{'WW}_T (k^2) |_{k^2=m_W^2} =  0 \ , \\\nonumber
Re \hat{\Sigma}^{ZZ}_T(m_Z^2) &=&  0 \,, \quad Re \hat{\Sigma}^{'ZZ}_T (k^2) |_{k^2=m_Z^2} =  0 \ , \\\nonumber
Re \hat{\Sigma}^{\gamma\gamma}_T(0) &=& 0 \,, \quad Re \hat{\Sigma}^{'\gamma\gamma}_T (k^2) |_{k^2=0} =  0\ , \\\nonumber
Re \hat{\Sigma}^{Z\gamma}_T(m_Z^2) &=&  0 \,,\quad  Re \hat{\Sigma}^{\gamma Z}_T(0) =  0\ . \nonumber
\end{eqnarray} 
The function $\widetilde{Re}$ takes the real part of the loop integrals only and it does not remove the imaginary parts arising 
from the various couplings of the theory (e.g. from complex CKM matrix elements); 
$\hat{\Sigma}'$ instead is a short-hand notation for $\hat{\Sigma}'(k^2)=\partial \hat{\Sigma}(k^2)/\partial k^2$. 
The definitions of $\hat{\Sigma}(k^2)$ are \cite{ultimo2Robens,Denner}\footnote{ Differently from the approach of Ref.\cite{ultimo2Robens}, where $Re \hat{\Sigma}^{HS}(k^2)$ is expressed in terms of the mixed mass
counterterm $\delta m^2_{HS}$, we used the mixing angle counterterm $\delta \alpha$. The two approaches are related as follows: $\delta m^2_{HS}=(m_S^2 - m_H^2)\delta\alpha$.}:
\begin{align}
\label{sigmaW}
\widetilde{Re} \hat{\Sigma}^{WW}_T(p^2) =& \ \widetilde{Re} \Sigma^{WW}_T(p^2) +\delta Z_W(p^2 - m_W^2) - \delta m_W^2\,, \\
Re \hat{\Sigma}^{ZZ}_T(p^2) =& \ Re \Sigma^{ZZ}_T(p^2) +\delta Z_Z(p^2 - m_Z^2) - \delta m_Z^2\,, \\
Re \hat{\Sigma}^{\gamma \gamma}_T(p^2) =& \ Re \Sigma^{\gamma \gamma}_T(p^2) + p^2\delta Z_{\gamma} \,, \\
Re \hat{\Sigma}^{\gamma Z}_T(p^2) =& \ Re \Sigma^{\gamma Z}_T(p^2) +\frac{1}{2}\delta Z_{\gamma Z}(2p^2 - m_Z^2) + m_Z^2 \delta \theta_W \,, \\
Re \hat{\Sigma}^{HH}(p^2) =& \ Re \Sigma^{HH}(p^2) +\delta Z_{H}(p^2 - m_{H}^2) - \delta m_{H}^2\,, \\
\label{sigmaS}
Re \hat{\Sigma}^{SS}(p^2) =& \ Re \Sigma^{SS}(p^2) +\delta Z_{S}(p^2 - m_{S}^2) - \delta m_{S}^2\,, \\
Re \hat{\Sigma}^{HS}(p^2) =& \ Re \hat{\Sigma}^{SH}(p^2) \ = \ Re \Sigma^{HS}(p^2) + (m_H^2 - m_S^2)\delta \alpha \, + \nonumber \\ 
\label{condition2}
&\ \ \ \ \ \ \ \ \ +\left[\frac{\delta Z_{HS}}{2}(p^2 - m_H^2)+\frac{\delta Z_{SH}}{2}(p^2 - m_S^2)\right]\,,
\end{align} 
where we can put (for example, see eq.(38) of \cite{kanemura1}):
\begin{align}
\label{ZW}
\delta Z_W \ &= \ \delta Z_{\gamma} + \frac{c_W}{s_W}\delta Z_{\gamma Z}\,.
\end{align}
Finally, from the renormalization conditions above and eqs.(\ref{sigmaW}-\ref{condition2}) we can extract the counterterms:
\begin{align}
\label{count0}
\delta m_H^2 &= \ Re \Sigma^{HH}(m_H^2) \ , \ \ \ \ \ \  \delta Z_H = \ -Re \Sigma^{'HH}(k^2) |_{k^2=m_H^2} \ , \\
\label{count1}
\delta m_S^2 &= \ Re \Sigma^{SS}(m_S^2) \ , \ \ \ \ \ \ \,\,\,\, \delta Z_S = \ -Re \Sigma^{'SS}(k^2) |_{k^2=m_S^2} \ , \\
\label{count2}
\delta m_W^2 &= \ \widetilde{Re} \Sigma^{WW}_T(m_W^2) \ , \ \ \ \,\,\delta Z_W = \ -\widetilde{Re} \Sigma^{'WW}_T(k^2) |_{k^2=m_W^2} \ , \\
\label{count3}
\delta m_Z^2 &= \ Re \Sigma^{ZZ}_T(m_Z^2) \ , \ \ \ \ \ \ \ \delta Z_Z = \ -Re \Sigma^{'ZZ}_T(k^2) |_{k^2=m_Z^2} \ , \\ 
\label{count4}
\delta Z_{\gamma Z} = \ 2 &Re\frac{\Sigma^{\gamma Z}_T(0)}{m_Z^2} + \frac{\delta s_W^2}{s_W c_W} \ \ , \ \ \delta Z_{Z\gamma} = \ -2 Re\frac{\Sigma^{\gamma Z}_T(m_Z^2)}{m_Z^2} - \frac{\delta s_W^2}{s_W c_W}\,, \\
& \ \ \ \ \ \ \ \ \ \ \ \ \ \,\, \delta Z_{\gamma} = -Re \Sigma^{'\gamma\gamma}_T(k^2) |_{k^2=0} \,.
%\delta Z_{HS} &= \ \delta Z_{SH} = \ \frac{1}{(m_S^2-m_H^2)}\left[Re\Sigma_{HS}(m_H^2)-Re\Sigma_{HS}(m_S^2)\right], \\
%&\delta \alpha = \ \frac{1}{2(m_S^2-m_H^2)}\left[Re\Sigma_{HS}(m_H^2)+Re\Sigma_{HS}(m_S^2)\right]\,,
\end{align} 
The derived quantities can be expressed in terms of the counterterms derived above, such as:
\begin{align}
\label{count5}
\delta &s_W^2 = - \delta c_W^2 = c_W^2 \bigg(\frac{\delta m_Z^2}{m_Z^2} - \frac{\delta m_W^2}{m_W^2} \bigg) \ \ , \ \ \delta Z_e = -\frac{1}{2}\delta Z_{\gamma} + \frac{s_W}{c_W}\frac{Re\Sigma^{\gamma Z}_T(0)}{m_Z^2} \,. 
%\  \delta s_W = \frac{1}{2}\frac{\delta s_W^2}{s_W}.
\end{align} 
Now, to fix the non-diagonal scalar field renormalization $\delta Z_{HS}$, we consider the improved on-shell (iOS) renormalization scheme \cite{ultimo2Robens} which requires that loop-induced $S-H$ or $H-S$ transitions 
vanish for on-shell external scalar states:
\begin{align}
\label{iOScond}
 Re \hat{\Sigma}^{HS}(p^2)\bigr|_{p^2=m_H^2}\,= \,0 \ \ \ , \ \ \ Re \hat{\Sigma}^{HS}(p^2)\bigr|_{p^2=m_S^2}\,=\,0\,.
\end{align}
The iOS definitions of $\delta Z_{HS\,,SH}^{\rm \, ios}$ (we indicate with the superscript "{\rm ios}"
the counterterm of the mixing scalar sector arising from the iOS renormalization conditions) are determined using eq.(\ref{condition2}) and eq.(\ref{iOScond}).
These equations lead to \cite{ultimo2Robens},
\begin{align}
\label{improvedEQ}
\frac{\delta Z_{HS}^{\rm \, ios}}{2}\,&=\, \frac{Re\Sigma^{HS}(m_S^2) }{m_H^2-m_S^2} + \delta\alpha^{\rm ios} \ \ \,, \ \ \frac{\delta Z_{SH}^{\rm \, ios}}{2}\,=\,
\frac{Re\Sigma^{HS}(m_H^2) }{m_S^2-m_H^2} - \delta\alpha^{\rm ios} \,.
\end{align}
On the other hand, the mixed mass (mixing angle) counterterm is defined in the following way \cite{ultimo2Robens}:
\begin{align}
\label{improvedEQ2}
\delta m^{2\,\,{\rm ios}}_{HS} \,=\, (m_S^2 - m_H^2)\,\delta \alpha^{\rm \, ios}\,=\, Re\Sigma^{HS}(p^{*2})\bigg|_{p^{*2}=\frac{m_H^2+m_S^2}{2}}\,,
\end{align}
where $p^{*2}$ is fixed to the average of the squared masses.
%and the relation between $(\delta m_{HS}^2)^{\rm \, ios}$ and $\delta \alpha^{\rm \, ios}$ is pointed out (see eq.(\ref{mhsalp}))
The reason for such a choice of $p^{*2}$ lies on the fact that the mixed scalar self-energy at $p^{*2}$ is independent on the gauge-fixing scheme. The gauge independence of the iOS scheme 
is also discussed in \cite{Espinosa:2002cd} where it is shown how the mixed scalar self energy at $p^{*2}$ coincides with the gauge invariant part of the same quantity
obtained through the so-called {\it pinch technique}, which generally allows the construction of off-shell Green's functions in non-Abelian gauge \cite{Binosi:2009qm} or extended scalar \cite{kanemura4} theories that 
are independent of the gauge-fixing parameter.\\
% 
% For example, in \cite{kanemura4} the authors calculate the NLO corrections to the Higgs boson couplings based on the OS renormalization scheme by using 
% the pinch technique to remove the gauge dependence. The cancellation of the gauge dependence is also directly proven in the Higgs boson two-point functions computed in the linear $R_\xi$ gauge by adding "pinch-terms" which are extracted from
% vertex corrections and box diagrams of a fermionic scattering process of the type $\bar{f}f\to \bar{f}f$.\\
The analytic results presented in this paper have been obtained using {\sc FeynRules} \cite{Alloul:2013bka} to generate the Feynman rules for the SSM model. 
% The counterterm structure is determined and implemented by hand. 
All amplitudes are computed with {{\sc FeynArts} \cite{Hahn:2000kx}} while their analytical processing was done with {{\sc FormCalc} \cite{Hahn:2000kx}}. The outputs,  written in terms of standard loop integrals,
have been evaluated with the help of {{\sc Package-X} \cite{Patel:2015tea}}.
%further reduced via Passarino-Veltman decomposition \cite{PassVelt1} and
\section{Renormalization of the $SVV$ vertex in the SSM}
\label{SVVvertex}
In this section we apply the renormalization procedure to the vertex of the scalar field  $S$ with two gauge bosons.
The related Feynman diagrams are reported in App.\ref{FeynDiag}, where we only show the contributions due to the insertion of the $S$ field in the loops because loops with the SM fields 
are as those quoted in \cite{kniehl1} where the external Higgs leg can be replaced with the new scalar singlet.
The bare $\mathcal{V}_0$ and one-loop corrections to the  $SVV$ vertex can be put in the following form \cite{kniehl1}:
\begin{align}
\label{vertex1loop}
\mathcal{V}^{\mu\nu} = \ \mathcal{V}_0^{\mu\nu} + \rho_V T^{\mu\nu} \ ,
\end{align} 
where $\rho_V$  has been defined in eq.(\ref{rho}) and $\mathcal{V}_{0}^{\mu\nu}=\rho_V g^{\mu\nu} $.
The generic expansion of $T^{\mu\nu}$ in terms of 2-index tensors is given by \cite{kniehl1}:
\begin{align}
\label{tensor}
T_V^{\mu\nu} = \mathcal{A}_V p^\mu p^\nu + \mathcal{B}_V q^\mu q^\nu + \mathcal{C}_V p^\mu q^\nu + \mathcal{D}_V q^\mu p^\nu + \mathcal{E}_V g^{\mu\nu} + i \mathcal{F}_V \epsilon^{\mu\nu\rho\sigma}p_{\rho}q_{\sigma}\,,
\end{align} 
where $p$ and $q$ are the four-vectors of the external gauge bosons. The  coefficients $\mathcal{A}_V,...,\mathcal{F}_V$ have to be ultra-violet (UV) finite whereas the term proportional to the antisymmetric tensor $\epsilon^{\mu\nu\rho\sigma}$
vanishes due to the charge conjugation invariance for external Z bosons and also if the gauge bosons are on the mass-shell.
We decide to set the external squared momenta $[k^2,p^2,q^2]$ in the $S(k)V(p,a)V(q,b)$ vertex as $\left[m_S^2,m_V^2,m_V^2\right]$, respectively. We take physical gauge bosons, so that only the coefficients
$\mathcal{D}_V$ and $\mathcal{E}_V$ become relevant.
Since the counterterms arising from the quantities in eq.(\ref{rho}) are included in the coefficient $\mathcal{E}_V$, we put it in the form $\mathcal{E}_V = \delta \rho_V^{CT} + \delta {\cal V}^{\mathcal{E}}_V$, 
where the symbol $\delta {\cal V}$ is used to indicate the three point function contributions at the one loop level (whose cumbersome expressions are not reported here) and $\delta\rho_V^{CT}$ are given by:
\begin{align}
\label{rhoctW}
\delta\rho_W &= \frac{\delta m_W^2}{2m_W^2} - \frac{\delta s_W^2}{2s_W^2} + \delta Z_W + \delta Z_e
+ \frac{\delta Z_S}{2} + \frac{c_\alpha}{s_\alpha}\bigg(\frac{\delta Z^{\rm ios}_{HS}}{2} -\delta \alpha^{\rm ios} \bigg)+ \frac{\delta s_\alpha^{\rm ios}}{s_\alpha}\,, \\
\label{rhoctZ}
\delta\rho_Z &= \frac{\delta m_Z^2}{m_Z^2} - \frac{\delta m_W^2}{2m_W^2} - \frac{\delta s_W^2}{2s_W^2} + \delta Z_Z + \delta Z_e 
+ \frac{\delta Z_S}{2} + \frac{c_\alpha}{s_\alpha}\bigg(\frac{\delta Z^{\rm ios}_{HS}}{2} -\delta \alpha^{\rm ios} \bigg)+ \frac{\delta s_\alpha^{\rm ios}}{s_\alpha}\,.
\end{align}
Notice that we obtain $\delta\rho_V$ independent from the mixing angle counterterm using the following substitution: $\delta s_\alpha^{\rm ios}= c_\alpha \delta \alpha^{\rm ios}$.
The other counterterms entering the previous expressions have been listed in eqs.(\ref{count1}-\ref{count5}) and eq.(\ref{improvedEQ}). 
In the following we will work in the 
\textit{modified on-shell mass scheme} (MOMS), in which the electric charge dependence in the coupling 
is replaced by the  Fermi constant $G_F$  via 
\begin{align}
\label{GF}
\frac{G_F}{\sqrt{2}} \ = \ \frac{e^2}{8 s_W^2 m_W^2}\frac{1}{1-\Delta r}\,.
\end{align}
This prevents the appearance of ambiguities associated with
the definition of the light-quark masses and mass singularities due to light fermions in $\ln(m_Z^2/m_f^2)$ terms in $\delta Z_e$
(more precisely, these logarithms appear in $\delta Z_\gamma$ which is part of $\delta Z_e$). 
% to show up in $\delta\rho_V^{CT}$.
%\footnote{This also implies that W-boson mass $m_W$ becomes: $$ m_W^2=\frac{m_Z^2}{2}\left(1 + \sqrt{1-2\pi\sqrt{2}\alpha_{em}/G_F m_Z^2} \ \right).$$}. 
The factor $\Delta r$ represents finite corrections to $G_F$; these are well known and up to ${\cal{O}}(\alpha^2_{em})$ are given by \cite{kniehl1,Sirlin:1980nh} \footnote{
In the SSM, the new contributions to $\Delta r$ generate a maximum variation of ${\cal O}(0.1)\%$ for $|s_\alpha|\sim 0.2$ and $m_S\geq 200$ GeV \cite{deltaR}.}: 
\begin{align}
\label{deltar}
\Delta r \ = \ \frac{\widetilde{Re} \hat{\Sigma}^{WW}_T(0)}{m_W^2} + \frac{\alpha_{em}}{4\pi s_W^2}\left[\left( \frac{7}{2 s_W^2}-2\right)\ln [c_W^2] + 6 \right],
\end{align}
where $\widetilde{Re} \hat{\Sigma}^{WW}_T(p^2)$ is the renormalized transverse self-energy of the W boson at momentum transfer $p$ defined in eq.(\ref{sigmaW}) and the last term is due to the vertex-box loop corrections
in the muon decay process.
The use of $G_F$ instead of the electric charge amounts to shift  $\delta \rho_V^{CT} \to \delta {\rho_V^{CT}}\sp{\prime} = \delta \rho_V^{CT} - \Delta r/2 $;
if in $\Delta r$ we use the form of $\delta Z_W$ as given in eq.(\ref{ZW}), the cancellation of the $\delta Z_{\gamma}$ in the final counterterm contributions is 
guaranteed and no problem arises from the light fermion loop contributions.
So finally we get:
\begin{align}
\label{deltalamS}
\delta{\rho_V}\sp{\prime} &= \ \delta Z_V + \frac{\delta Z_S}{2} + \frac{c_\alpha}{s_\alpha}\frac{\delta Z^{\rm ios}_{HS}}{2} + \frac{\delta m_V^2}{m_V^2} 
-\frac{\widetilde{Re}\Sigma_T^{WW}(0)}{2 m_W^2} \ + \nonumber \\
& \ \ \ \ \ \ \ \ \ \ \ \ \ \ \ \ + \frac{Re\Sigma^{AZ}_T(0)}{s_W c_W\,m_Z^2} -\frac{\alpha_{em}}{8\pi s_W^2}\left[\left( \frac{7}{2 s_W^2}-2\right)\ln [c_W^2] + 6 \right]\,.
\end{align} 
The coefficient $\mathcal{E}_V$ is UV-finite both for Z and W external boson pairs, as it can be explicitly verified 
from the expressions of the bosonic and fermionic divergent parts quoted in Tab.\ref{DIVE} for all the counterterms (which are divided by a common factor $g^2 / (16 \pi^2\, \epsilon)$).
\begin{table}[h!]
\begin{center}
\begin{tabular}{c  c  c  c  c  c}
\toprule
\toprule
& $\ \ \ \ $ & $ \mathcal{E}_W $ & $ \mathcal{E}_Z $ & $\textrm{UV}_{\textrm{bosonic}}$ & $\textrm{UV}_{\textrm{fermionic}}$ \\
\midrule
\vspace{1.5mm}
& $\delta Z_W$ & $\text{\large \cmark}$& $\text{\large \xmark}$& $19/6$ & $-4$ \\
\vspace{1.5mm}
& $\delta Z_Z$ & $\text{\large \xmark}$& $\text{\large \cmark}$& $\frac{-1+2c_W^2+18c_W^4}{6c_W^2} $ & $\frac{-20+40c_W^2-32c_W^4}{3c_W^2} $ \\
\vspace{1.5mm}
& $\delta Z_S/2$ & $\text{\large \cmark}$& $\text{\large \cmark}$& $\frac{s_\alpha^2(2c_W^2+1)}{4c_W^2} $ & $-\frac{s_\alpha^2\sum_f N_c m_f^2}{4m_W^2}$ \\
\vspace{1.5mm}
& $c_\alpha\delta Z^{\rm ios}_{HS}/2 s_\alpha$ & $\text{\large \cmark}$& $\text{\large \cmark}$& $\frac{c_\alpha^2(2c_W^2+1)}{4c_W^2} $ & $-\frac{c_\alpha^2\sum_f N_c m_f^2}{4m_W^2}$ \\
\vspace{1.5mm}
& $\delta m_W^2/m_W^2$ & $\text{\large \cmark}$& $\text{\large \xmark}$& $\frac{6-31c_W^2}{6c_W^2}$ & $4 -\frac{\sum_f N_c m_f^2}{2 m_W^2}$ \\
\vspace{1.5mm}
& $\delta m_Z^2/m_Z^2$ & $\text{\large \xmark}$& $\text{\large \cmark}$& $\frac{7+10c_W^2-42c_W^4}{6c_W^2}$ & $\frac{20-40c_W^2+32c_W^4}{3c_W^2} -\frac{\sum_f N_c m_f^2}{2m_W^2}$ \\
\vspace{1.5mm}
& $-\widetilde{Re}\Sigma_T^{WW}(0)/2m_W^2$ & $\text{\large \cmark}$& $\text{\large \cmark}$& $\frac{2c_W^2 -1}{2c_W^2}$ & $\frac{\sum_f N_c m_f^2}{4m_W^2}$ \\
\vspace{1.5mm}
& $Re\Sigma^{AZ}_T(0)/s_W c_W m_Z^2$ & $\text{\large \cmark}$& $\text{\large \cmark}$& $-2$ & 0 \\
\vspace{1.5mm}
& $\delta {\cal V}^{\mathcal{E}}_W$ & $\text{\large \cmark}$& $\text{\large \xmark}$& $\frac{-3+10c_W^2}{4c_W^2}$ & $\frac{\sum_f N_c m_f^2}{2m_W^2}$ \\
\vspace{1.5mm}
& $\delta {\cal V}^{\mathcal{E}}_Z$ & $\text{\large \xmark}$& $\text{\large \cmark}$& $\frac{-3-6c_W^2+16c_W^4}{4c_W^2}$ & $\frac{\sum_f N_c m_f^2}{2m_W^2}$ \\
\bottomrule
\bottomrule
\end{tabular}
\caption{\it Coefficients of the bosonic and fermionic  UV divergent parts of the relevant counterterms (which are divided by the common factor $g^2 / (16 \pi^2\, \epsilon)$).
The symbol $\text{\large \cmark}$($\text{\large \xmark}$) indicates that the corresponding counterterm is present (absent) in $\mathcal{E}_{W,Z}$.}
\label{DIVE}
\end{center}
\end{table}
Regarding the finite parts, we know that the $S$ field gives negligible contributions to the corrections of the muon decay and since there is no $S$ field 
dependence in $\Sigma^{AZ}_T(0)$ \cite{deltaR}, the new scalar contributions only affect the bosonic parts of
$\widetilde{Re}\Sigma_T^{WW}(0)$, $\delta m_Z^2$, $\delta Z_Z$, $\delta Z_S$, $\delta Z^{\rm ios}_{HS}$ and $\delta {\cal V}^{\mathcal{E}}_V$. The fermionic contributions of 
$\widetilde{Re}\Sigma_T^{WW}(0)$, $\delta m_Z^2$, $\delta Z_Z$ and $\delta {\cal V}^{\mathcal{E}}_V$ are identical to those associated to the $HVV$ vertex 
in the SM; in addition, their contributions to $\delta Z_S$ and $\delta Z^{\rm ios}_{HS}$ can be determined from the fermion loop terms in the SM
$\delta Z_H$ expression but now multiplied by $s_\alpha^2$ and $c_\alpha s_\alpha$ and with external momenta fixed to $m_S^2$ (the definition of $\delta Z^{\rm ios}_{HS}$ also contains the mixed two-point
function with external momenta fixed to $(m_H^2+m_S^2)/2$).\\
Notice that the renormalization of the  vertex $S {W}^+ W^{-}$ is more complicated than the $SZZ$  vertex since contributions 
due to the photons in the loop integrals, which are plagued by infrared (IR) singularities when the W bosons are on-shell, must be taken into account. 
The IR-cancellation is obtained considering soft-photon bremsstrahlung contributions \cite{kniehl2} which, for the model under discussion, are shown in Fig.(\ref{Brem}).
\section{The decay $S \to VV$ in the SSM}
\label{SVVdecay}
\begin{figure}[b!]
\centering
\includegraphics[scale=0.5]{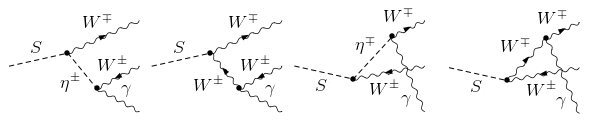}
\caption{\it Feynman diagrams of the photon bremsstrahlung associated to the first-order radiative corrected $SW^+W^-(\gamma)$ vertex.}
\label{Brem}
\end{figure}

The decay rate of the scalar $S$ into two physical gauge bosons gets contributions from longitudinally 
($L$) and transversally ($\pm\pm$) polarized gauge bosons.
Using the LO amplitude of eq.(\ref{couszz}), a straightforward computation of the decay width gives,
\begin{equation}
\Gamma^{\textrm{LO}}_{VV} \ = \ \frac{G_F}{16\sqrt{2}\pi} m_S^3 s_\alpha^2 (1+\delta_V) \sqrt{1-4x_V} \left(1- 4x_V +12x_V^2\right)\,,
\end{equation}
where $x_V=m_V^2/m_S^2$, $\Gamma_{ii}=\Gamma(S\to ii)$ and $\delta_V =0,1$ for $V=Z,W^\pm$ respectively. The longitudinally and transversally polarized gauge boson contributions to $\Gamma^{\textrm{LO}}_{VV}$ are given by \cite{kniehl1,kniehl2}:
\begin{align}
(\Gamma^{\textrm{LO}}_{VV})^{\pm L} \ =& \ (\Gamma^{\textrm{LO}}_{VV})^{\pm\mp} \ = \ 0\,, \\
\label{tras}
(\Gamma^{\textrm{LO}}_{VV})^{\pm\pm} \ =& \ \frac{G_F}{16\sqrt{2}\pi} m_S^3 s_\alpha^2 (1+\delta_V) \sqrt{1-4x_V} \times (4x_V^2)\,, \\
\label{long}
(\Gamma^{\textrm{LO}}_{VV})^{LL} \ =& \ \frac{G_F}{16\sqrt{2}\pi} m_S^3 s_\alpha^2 (1+\delta_V) \sqrt{1-4x_V} \times \left(1- 4x_V +4x_V^2\right)\,.
\end{align}
As a consequence of eq.(\ref{tensor}) and of the polarization conditions ($\epsilon_\mu(p) \cdot p^\mu = \epsilon_\nu(q) \cdot q^\nu = 0$), the physical amplitude is reduced to:
\begin{equation}
\label{vertex1loopS}
\mathcal{V}^{\mu\nu}\epsilon_\mu(p)\epsilon_\nu(q)\ = \ \rho_V [g^{\mu\nu}(1+\mathcal{E}_V) + q^\mu p^\nu \mathcal{D}_V]\epsilon_\mu(p)\epsilon_\nu(q)\,.
\end{equation}
As mentioned above, the radiative corrections to $S \to W^+W^-$ are guaranteed to be IR-finite if we include soft-photon bremsstrahlung contributions which, for the model under discussion, are shown in Fig.(\ref{Brem}).
We call the photon momenta as $q_\gamma$ whose maximum value is $q_\gamma^{\textrm{max}}=m_S(1-4x_W)/2$. 
% To regularize the IR-divergences it is necessary to assign a virtual mass $m_\gamma$ to the photon which works as an infrared regulator. 
% Tipically, a bremsstrahlung photon can be a soft or a 
% hard photon. Differently from hard photons which are detected in the final state, the soft photons have tipical energies smaller than the energy threshold
% of the experiment and they are not detected. 
To set an ideal boundary between the soft and hard region, we introduce a cutoff $\Lambda_\gamma$ in such a way that the soft region corresponds to 
$m_\gamma \leq q_\gamma \leq \Lambda_\gamma$ while the hard region to $\Lambda_\gamma \leq q_\gamma \leq q_\gamma^{\textrm{max}}$, where $m_\gamma$ is a fake mass assigned to the photon.
The total photon-bremsstrahlung decay rate 
is then given by the sum of the soft and hard contributions:
\begin{align}
\label{gbrem}
\Gamma^{\textrm{brem}}_{WW} \ =\Gamma^{\textrm{soft}}_{WW}+\Gamma^{\textrm{hard}}_{WW} \ = \Gamma^{\textrm{LO}}_{WW}(\delta^{\textrm{soft}}_W+\delta^{\textrm{hard}}_{W}),
\end{align}
where the correction factors $\delta^{\textrm{soft}}_{W}$ and $\delta^{\textrm{hard}}_{W}$ are reported in App.\ref{BREM} and are extracted from \cite{kniehl2}.
The $m_\gamma$ and $\Lambda_\gamma$ dependences show up in $\delta Z_W$, $\delta {\cal V}^{\mathcal{E}}_W$, $\delta^{\textrm{soft}}_{W}$ and $\delta^{\textrm{hard}}_{W}$, as detailed in Tab.\ref{IR}.
\begin{table}[h!]
\begin{center}
\begin{tabular}{c  c  c  c}
\toprule
\toprule
& $\ \ \ \ $& $m_\gamma$ (IR regulator) & $\Lambda_\gamma$ (IR cutoff)\\
\midrule
& $\delta Z_W$ & $\frac{\alpha_{em}}{2\pi}\ln\bigg(\frac{m_W^2}{m_\gamma^2}\bigg)$ & ----- \\
& $\delta {\cal V}^{\mathcal{E}}_W$ & $\frac{\alpha_{em}}{2\pi}\left[\mathcal{G}(r)+1\right]\ln\bigg(\frac{m_\gamma^2}{m_W^2}\bigg)$ & ----- \\
& $\delta^{\textrm{soft}}_{W}$ & $\frac{\alpha_{em}}{\pi}\mathcal{G}(r)\ln \bigg(\frac{m_W^2}{m_\gamma^2}\bigg)$ & $\frac{\alpha_{em}}{\pi}\mathcal{G}(r)\ln \bigg(\frac{4\Lambda_\gamma^2}{m_W^2}\bigg)$\\
& $\delta^{\textrm{hard}}_{W}$ & ----- & $\frac{\alpha_{em}}{\pi}\mathcal{G}(r)\ln \bigg(\frac{m_S^2}{4\Lambda_\gamma^2}\bigg)$ \\
\bottomrule
\bottomrule
\end{tabular}
\caption{\small {\it IR-dependence on $m_\gamma$ and $\Lambda_\gamma$ in} $\delta Z_W$, $\delta {\cal V}^{\mathcal{E}}_W$, $\delta^{\textrm{soft}}_{W}$ {\it and} $ \, \delta^{\textrm{hard}}_{W}$.}
\label{IR}
\end{center}
\end{table}
Here $r = m_S^2/4m_W^2$ and  the function $\mathcal{G}(r)$ is defined in App.\ref{BREM}.\\
The NLO total decay width, which we call $\Gamma^{\textrm{NLO}}_{VV}$, is given by the sum of $\Gamma_{VV}^{\pm\pm}$ and $\Gamma_{VV}^{L}$ 
at the one-loop order and $\Gamma^{\textrm{brem}}_{WW}$ of eq.(\ref{gbrem}). So we have,
\begin{align}
%(\Gamma^{\textrm{NLO}}_V)^{L} + 2(\Gamma^{\textrm{NLO}}_V)^{\pm\pm} + \delta_V \Gamma_{\textrm{brem}} =
\label{gammastot}
& \ \ \ \ \ \ \ \Gamma^{\textrm{NLO}}_{VV} = \ \frac{G_F}{16\sqrt{2}\pi} m_S^3 s_\alpha^2 (1 +\delta_V)\sqrt{1-4x_V}\left(1- 4x_V +12x_V^2\right) \times  \, \nonumber \\
&\times\bigg\{1 + 2\bigg[\delta{\rho_V}\sp{\prime} + \delta {\cal V}^{\mathcal{E}}_V + \frac{m_S^2}{2}\left(\frac{1-6x_V + 8x_V^2}{1- 4x_V +12x_V^2}\right)\delta {\cal V}^{\mathcal{D}}_V \bigg] \bigg\} + \delta_V \Gamma_{WW}^{\textrm{brem}}\,,
\end{align}
where $\delta {\cal V}^{\mathcal{D}}_V$ are the corrections from the coefficient $\mathcal{D}_V$.
% and $\delta V_{\rm MF}$ represents the wave-function correction discussed in eq.(\ref{finiteWF}) for the minimal field renormalization scheme\footnote{Notice that $\delta V_{\rm MF}=0$ at $\mu_R=m_S$.}:
% \begin{align}
% \delta V_{\rm MF}\,=\, \frac{c_\alpha}{s_\alpha}\hat{Z}_{SH} \ = \ \frac{c_\alpha}{s_\alpha}\frac{\hat{\Sigma}^{SH}(m_S^2)}{(m_H^2-m_S^2)}\,.
% \end{align}
Now, two comments are in order:\\
i) using the $m_\gamma$-dependent contributions, reported in Tab.\ref{IR}, we can verify the cancellation of the IR-divergences:
\begin{align}
\label{soft}
\{\Gamma^{\textrm{NLO}}_{WW} \}_{\textrm{IR}} \propto \left [ 1 + \delta^{\textrm{soft}}_{W} + 2(\delta Z_W + \delta {\cal V}^{\mathcal{E}}_W)\right]
\propto \left[ 1 + \frac{\alpha_{em}}{\pi}\mathcal{G}(r)\ln\bigg(\frac{4\Lambda_\gamma^2}{m_W^2}\bigg) \right]\,;
\end{align}
ii) the combination of all terms in Tab.\ref{IR} is $\Lambda_\gamma$-independent at $\mathcal{O}(\alpha_{em})$.
% processes without a hard photon emission are also possible and encounter the classic infrared catastrophe
% for a sufficently small value of $\Lambda_\gamma$ \cite{IR}.
% This problem is solved using the exponential soft photon contributions (Yennie-Frautschi-Suura Theorem \cite{YFS}):
% \begin{align}
% 1 + \frac{\alpha_{em}}{\pi}\mathcal{G}(r)\ln\bigg(\frac{4\Lambda_\gamma^2}{m_W^2}\bigg) + \frac{1}{2}\bigg[\frac{\alpha_{em}}{\pi}\mathcal{G}(r)\ln\bigg(\frac{4\Lambda_\gamma^2}{m_W^2}\bigg)\bigg]^2+...
% =\bigg(\frac{4\Lambda_\gamma^2}{m_W^2}\bigg)^{\frac{\alpha_{em}}{\pi}\mathcal{G}(r)}.
% \end{align} 
% This replacement ensures the suppression of the soft photon bremsstrahlung probability for decreasing values of the cutoff $\Lambda_\gamma$. 
% Notice that after this substitution we still have a $\Lambda_\gamma$ dependence of order $\mathcal{O}(\alpha_{em}^2)$ and this implies a choice of $\Lambda_\gamma$ value. 
% For a reasonable value of $\Lambda_\gamma$, the variation of $\Gamma^{\textrm{NLO}}_W$ is negligible as compared to $\mathcal{O}(\alpha_{em})$ result. Therefore
% we have developed our work at $\mathcal{O}(\alpha_{em})$ where is not necessary to fix $\Lambda_\gamma$.
\subsection{Numerical Results}
\label{SVVnumer}
\begin{figure}[h!]
\centering
\includegraphics[scale=0.45]{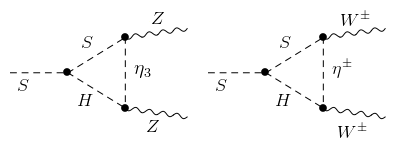}
\caption{\it Examples of the Feynman diagrams contributing to the mixing angle dependence of $\mathcal{R}^{\textrm{SSM}}_{VV}$.}
\label{diagSA}
\end{figure}

In the  evaluation of the corrections to the total decay rate we make use of the following quantity: 
\begin{equation}
\mathcal{R}^{\textrm{SSM}}_{VV}=[(\Gamma^{\textrm{NLO}}_{VV}/\Gamma^{\textrm{LO}}_{VV}) -1] \,.
\end{equation}
As a set of independent input parameters we choose $w$, $m_S$ and $\alpha$ and express at tree-level
$\lambda$, $\rho$ and $\kappa$ according to \cite{ultimo2Robens}:
\begin{align}
\label{systfull}
\lambda &= \ \frac{m_H^2 c_\alpha^2 + m_S^2 s_\alpha^2}{2 v^2} \ \ \ , \ \ \ \rho = \ \frac{m_S^2 c_\alpha^2 + m_H^2 s_\alpha^2}{2 w^2} \ \ \ , \ \ \ 
\kappa = \ \frac{(m_S^2 - m_H^2) s_{2\alpha}}{2 v w}.
\end{align}
The mass of the light scalar field is kept fixed to  $m_H = 125.09$ GeV. We then evaluate $\mathcal{R}^{\textrm{SSM}}_{VV}$ as a function of 
$s_\alpha$ for different values of $m_S$ and $w$. It has to be considered that the maximally allowed ranges for 
$|s_\alpha|$ depend on the assumed singlet mass \cite{status} and have been derived considering  
W boson mass measurement, electroweak precision observables  tested
via the oblique parameters S, T and U, perturbativity of the RG-evolved coupling $\lambda$ evaluated for
the exemplary choice $w/v=10$, perturbative unitarity, direct LHC searches and  Higgs
signal strength measurement \cite{ultimo2Robens}. On the other hand, perturbative unitarity poses a lower limit on 
the ratio $w/v$ which, again, depends on the singlet mass and $s_\alpha$. We summarize such informations on Tab.\ref{tab:parameter_table}, extracted from Table I of 
\cite{ultimo2Robens}, where we report the values of $m_S$ considered in our numerical analysis as well as the ranges of $|s_\alpha|$
and the corresponding $w_{min}$.
\begin{table}[h!]
\begin{center}
\begin{tabular}{c  c  c  c c}
\toprule
\toprule
& $m_S$ [GeV] & $|s_\alpha|$ & $w_{min}$ [GeV]\\
\midrule
 & 200 & [0.09,0.36] & 0.85 $v$ & \\ 
 & 300 & [0.067,0.31] & 1.25 $v$ & \\
 & 400 & [0.055,0.27] & 1.69 $v$ & \\
 & 500 & [0.046,0.24] & 2.13 $v$ & \\
 & 600 & [0.038,0.23] & 2.56 $v$ & \\
 & 700 & [0.031,0.21] & 3.03 $v$ & \\
 & 800 & [0.027,0.21] & 3.45 $v$ & \\
 & 900 & [0.022,0.19] & 3.85 $v$ & \\
 & 1000 & [0.018,0.17] & 4.34 $v$ & \\
\bottomrule
\bottomrule
\end{tabular}
\caption{ \it   Values of $m_S$ considered in our numerical analysis, the ranges of $|s_\alpha|$
and the corresponding $w_{min}$. Table extracted from Table I of \cite{ultimo2Robens}.}
\label{tab:parameter_table}
\end{center}
\end{table}
\begin{figure}[h!]
\centering
\includegraphics[scale=0.395]{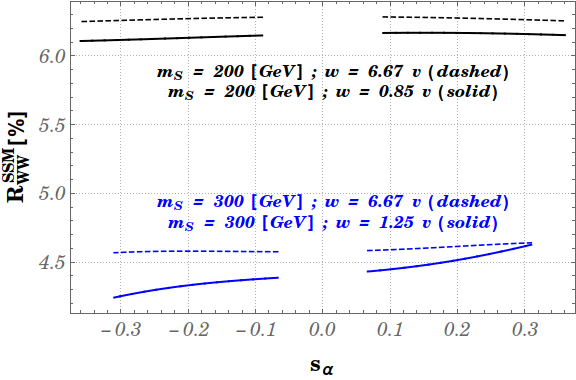}
\includegraphics[scale=0.405]{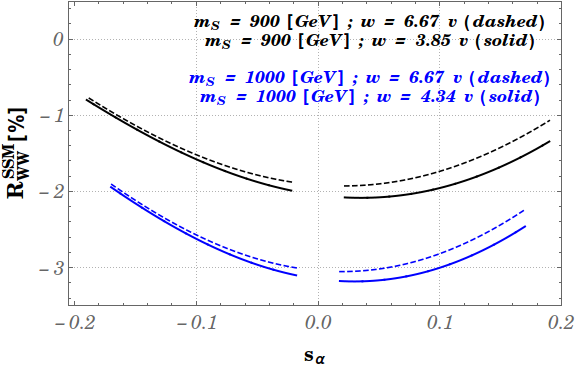}
\includegraphics[scale=0.395]{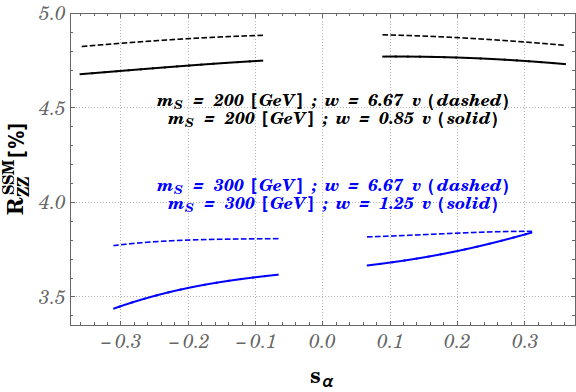}
\includegraphics[scale=0.405]{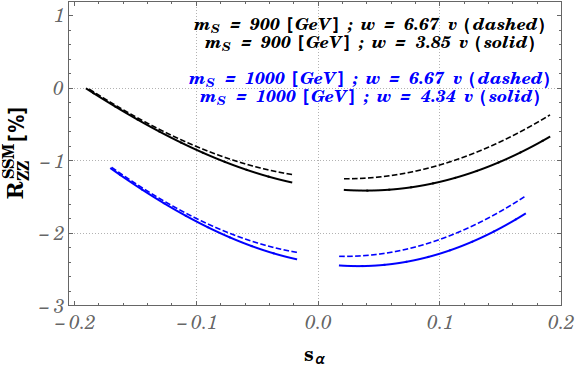}
\caption{\it $\mathcal{R}^{\textrm{SSM}}_{VV}$ as a function of $s_\alpha$, for different
values of $m_S$ (and the corresponding vev $w$). The range of $s_\alpha$ is the one deduced from Tab.\ref{tab:parameter_table}.
$\mathcal{R}^{\textrm{SSM}}_{WW}$ is computed with $q_\gamma= q_\gamma^{\textrm{max}}$.}
\label{RWZ}
\end{figure}
\\
\\
\\
\\
The numerical results for $\mathcal{R}^{\textrm{SSM}}_{VV}$ as a function of $s_\alpha$ are reported in Fig.\ref{RWZ}; in the upper panels we 
show the case $V=W$ (where for simplicity we fixed $q_\gamma= q_\gamma^{\textrm{max}}$) whereas in the lower ones $V=Z$.
For both cases we considered four possible values of $m_S$: a low mass region with $m_S=200, 300$ GeV (plots on the left) and a high mass region  with $m_S=900, 1000$ GeV (plots on the right). 
In order to roughly analyze the dependence on $w$, in the same plots we also show $\mathcal{R}^{\textrm{SSM}}_{VV}$ computed for two different values of the singlet vev $w$: the smallest
one (solid lines) is chosen according to the minimum reported in Tab.\ref{tab:parameter_table} while the largest is kept fixed at $w=6.67v$ (dashed lines), which is the value used
in \cite{ultimo2Robens} to determine the allowed intervals of $s_\alpha$ and, according to Tab.\ref{tab:parameter_table}, valid for every $m_S$.
First of all, we clearly see that the dependence on $w$ is not dramatic for every value of $m_S$. In fact, the absolute differences between the solid and the dashed lines amount to a maximum of $\sim 0.3 \%$
for both $\mathcal{R}^{\textrm{SSM}}_{ZZ}$ and $\mathcal{R}^{\textrm{SSM}}_{WW}$ when $m_S = 300$ GeV, $s_\alpha \sim -0.3$ and $m_S = 900$ GeV, $s_\alpha \sim 0.2$.
The reason for such a dependence is simply due to the fact that $\kappa$ (defined in eq.(\ref{systfull}) and entering in $\mathcal{R}^{\textrm{SSM}}_{VV}$) are inversely proportional to $w$ but grow with the mixing angle.\\
We also observe a different behavior with respect to $sign\left(s_\alpha\right)$; in particular, the ratios $\mathcal{R}^{\textrm{SSM}}_{VV}$ are weakly dependent on $s_\alpha$
in the low mass range while for the high mass region they increase or decrease proportionally to the mixing angle, especially for $s_\alpha < 0$. To be more 
quantitative: the two plots on the left-hand side of Fig.(\ref{RWZ}) show a maximum variation of each $\mathcal{R}^{\textrm{SSM}}_{VV}$ with respect to negative $s_\alpha$ of $\sim 0.2 \%$,
while for those on the right-hand side such a variation becomes $\sim 1.2 \%$ and $\sim 1.3 \%$ (for $V=W$ and $Z$, respectively). The high mass region also shows a variation with respect to the mixing angle 
in the region of $s_\alpha>0$ amounting to a maximum of $\sim 0.7 \%$ for both ratios.
The reason of this different behavior with respect to $sign\left(s_\alpha\right)$ has to be ascribed to those diagrams which contain odd powers of the coupling $\kappa$ which, according to eq.(\ref{systfull}), grows with $m_S^2$ 
and whose sign is only determined by $sign(s_\alpha)$ (for $m_S > m_H$, as it is the case in this paper). 
Typical Feynman diagrams with such a structure and that contribute to the mixing angle dependence of $\mathcal{R}^{\textrm{SSM}}_{VV}$ are depicted 
in Fig.(\ref{diagSA}). Neglecting the loop integrals for simplicity, 
the couplings evaluated  up to $\mathcal{O}(v^2/w^2)$ (we used the approximate expressions in eq.(\ref{limitdef})) are the following:
\begin{align}
(\textrm{SSH}) &\, \sim \, \kappa v \ \ \ \ , \ \ \ \ (\textrm{HV}\eta^{i}) \, \propto \, \frac{m_V}{v} \ \ \ \ , \ \ \ \ (\textrm{SV}\eta^{i}) \, \propto \, s_\alpha\frac{m_V}{v} \, \sim \, \frac{\kappa \, m_V}{2\rho w}\,, \nonumber 
\end{align}
which in turn imply an overall dependence given by:
\begin{align}
(\textrm{SSH}) \times (\textrm{SV}\eta^{i}) \times (\textrm{HV}\eta^{i}) \ \propto \ \frac{\kappa^2 \, m_V^2}{2\rho w} \ \sim \ \rho_V \, \kappa\,,
\end{align}
where $i=(3, \pm)$ for $V=Z,W^\pm$, respectively.
In the rest of this section we will scrutinize more in detail the dependence of $\mathcal{R}^{\textrm{SSM}}_{VV}$ on the singlet mass, separating the cases of positive and negative $s_\alpha$.\\
Notice that, increasing $m_S$, the range of the $|s_\alpha|$ taken into account is restricted to the following interval:
$|s_\alpha| \in [|s_\alpha|_{\textrm{min}}^{\textrm{m}_\textrm{S}\textrm{=200 GeV}}, |s_\alpha|_{\textrm{max}}^{\textrm{m}_\textrm{S}\textrm{=1000 GeV}}]=[0.09,0.17]$, which is valid for every choice of $m_S$ (see Tab.\ref{tab:parameter_table}). 
\subsection*{Case $s_\alpha > 0$}
\begin{figure}[h!]
\centering
\includegraphics[scale=0.51]{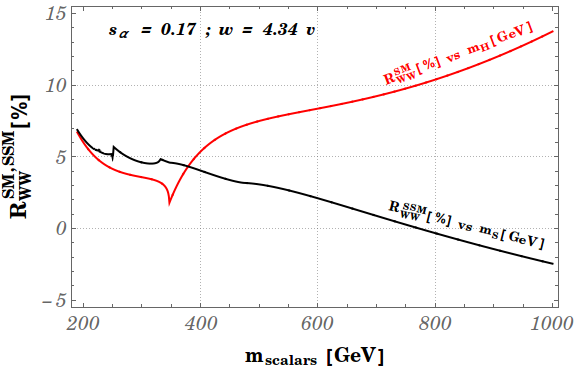}
\includegraphics[scale=0.51]{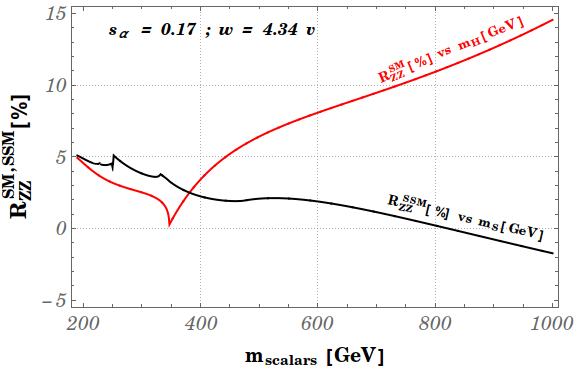}
\caption{\it $\mathcal{R}_{VV}^{\textrm{SM}}$ (red line) and $\mathcal{R}_{VV}^{\textrm{SSM}}$ (black line) as a function of the scalar mass  $m_{\rm{scalars}}$. In the case of the SSM, we fixed 
$(s_\alpha=0.17,\,w =4.34\, v)$.}
\label{SMvsSSM}
\end{figure}

\begin{figure}[h!]\hspace*{1.5cm}
\centering
\includegraphics[scale=0.5]{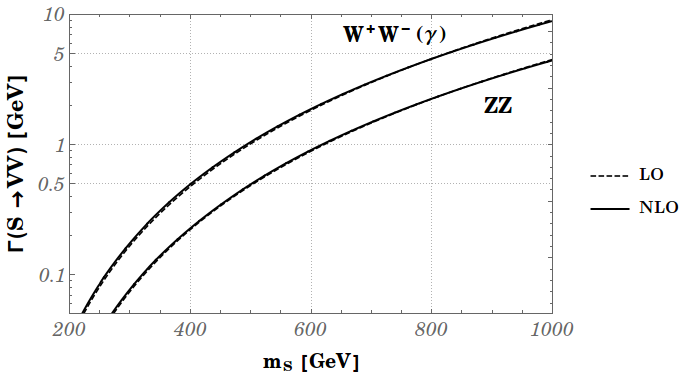}
\caption{\it Leading order (dashed line) and next-to-leading order (solid line) results for $\Gamma(S \to VV)$ ($s_\alpha=0.17$, $w=4.34 \,v$). 
In the case of $V=W$ the photon momenta is fixed to $q_\gamma=q_\gamma^{\textrm{max}}$.}
\label{gammap2}
\end{figure}

In Fig.(\ref{SMvsSSM}) we show the behavior of $\mathcal{R}^{\textrm{SSM}}_{VV}$ ($V=W$ on the left, $V=Z$ on the right) as a function of $m_S$ 
for a fixed $s_\alpha = 0.17$ and two different values of $w$, namely $w=(4.34, 6.67)\, v$.
For the sake of comparison, we also computed the same ratio $\mathcal{R}_{VV}^{\textrm{SM}}=[(\Gamma^{\textrm{NLO}}_{VV}/\Gamma^{\textrm{LO}}_{VV}) -1]$
in the SM (red line) leaving the Higgs mass as a free parameter 
(in practice, the SM with a heavy Higgs  \cite{kniehl1}). In the plots, on the common x-axis we use the label $m_{\rm{scalars}}$ to indicate either $m_H$ or $m_S$.
We observe two main differences.\\
\\
The first one is the finite peak at $m_S=2m_H$ (which corresponds to fix in the loop integrals: $k^2 = m_S^2 = 4m_H^2$) in $\mathcal{R}_{VV}^{\textrm{SSM}}$ due to the new coupling $SHH$ which is obviously absent in the SM.
The second is the different behavior in the heavy scalar mass region.
This is mainly due to the new scalar contributions arising from the coefficient $\mathcal{D}_V$ (see \cite{kniehl1} for an explicit evaluation in the SM). For example, setting 
the mass of the heavy scalars to $m_{\textrm{scalars}}=10^3$ GeV and considering $V=Z$, we have: 
$$\{ \delta {\cal V}^{\mathcal{D}}_Z\}^{\textrm{SM}}\sim \ (3.31 - 0.16 \ \lambda) \times 10^{-5} > 0\ ,$$
whose positivity is determined by the fact that $\lambda=m_{\textrm{scalars}}^2/2v^2$. Instead, in the case of the SSM, we get:
$$\{ \delta {\cal V}^{\mathcal{D}}_Z\}^{\textrm{SSM}}\sim \ (4.97 - 2.27 \ \lambda - 2.07 \ \rho - 13.95 \ \kappa)\times 10^{-5} < 0$$
because $\lambda, \rho$ and $\kappa$ are all positive parameters for $s_\alpha = 0.17$ and $w=(4.34, 6.67)\,v$, see eq.(\ref{systfull}).\\
The black curves show that $\mathcal{R}_{VV}^{\textrm{SSM}}$ exhibits positive (negative) corrections for values of the scalar mass below (above) $m_{\textrm{scalars}}\sim 750$ GeV.
Additional features concerning these curves are visible when $m_S \sim 230,\,330$ GeV (new kinks shortly below $m_H$ and $m_t$) and $m_S \sim 470$ GeV. These behaviors arise from the finite parts of the  
loop integrals ${\cal B}_0(k^2, M^2,M^2)$, with $M=m_Z,\,m_H,\,m_t$, which appear in $\delta\alpha^{\rm ios}$ defined at the singlet scalar momentum $k^{*2}=(m_S^2+m_H^2)/2$. As a consequence, 
the kinks are dictated by the condition,
\begin{align}
 k^{*2} \,=\,\frac{m_H^2+m_S^2}{2} \,=\,4M^2 \ \, \Longrightarrow \ \, m_S\,=\,\sqrt{8M^2 - m_H^2}\,\sim\, \{230,\,330,\,470\, {\rm GeV}\}\,.
\end{align}
%with $M=m_Z,\,m_H,\,m_t$, respectively.} 
Finally, in Fig.(\ref{gammap2}) we summarize our results for the decay widths $\Gamma(S \to Z Z )$ and $\Gamma(S \to W^+W^-(\gamma))$ as a function of $m_S$ for the selected values 
$s_\alpha=0.17$, $w=4.34 \,v$. As expected from our previous considerations, the NLO results (solid line) are above the LO behavior (dashed line) in the small mass region but 
becomes smaller in the region of larger masses.
\subsection*{Case $s_\alpha < 0$}
\begin{figure}[h!]
\centering
\includegraphics[scale=0.4]{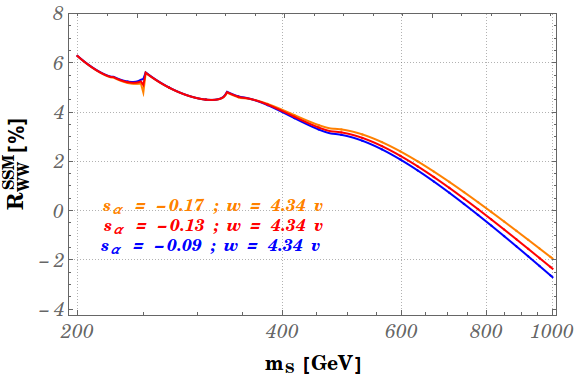}
\includegraphics[scale=0.4]{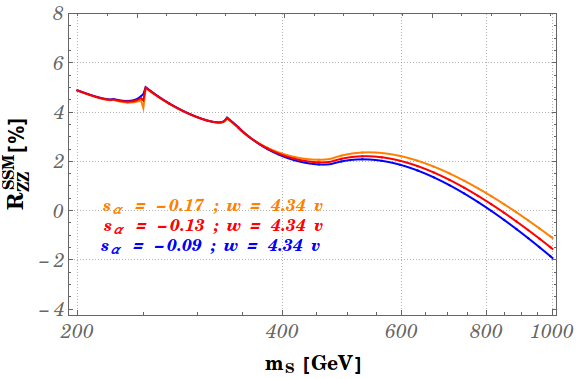}
\caption{\it Ratio $\mathcal{R}_{VV}^{\textrm{SSM}}$ as a function of the singlet mass $m_S$ for the three fixed values  $s_\alpha=-0.09, -0.13$ and $-0.17$.}
\label{RATIOgammasn}
\end{figure}

\begin{figure}[h!]\hspace*{2cm}
\centering
\includegraphics[scale=0.5]{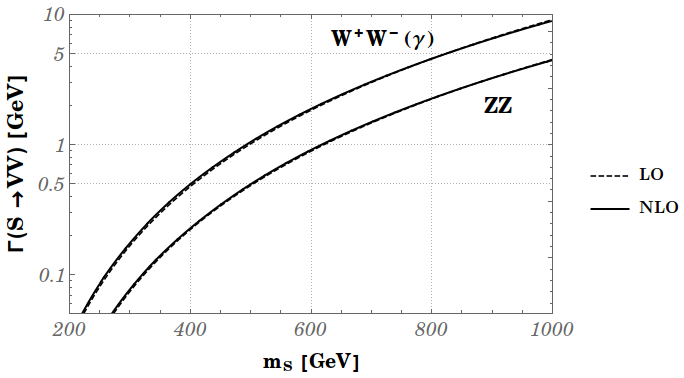}
\caption{\it The same as Fig.(\ref{gammap2}) but for $s_\alpha=-0.17$.}
\label{gamman2}
\end{figure}

As it was shown in Fig.(\ref{RWZ}), the ratio $\mathcal{R}_{VV}^{\textrm{SSM}}$ in the high mass region depends much more on the sign and on the assumed value of $\sin\alpha$ than the case of low-mass, and the variation with $sign(s_\alpha)$ 
is more evident when $s_\alpha <0$. To study more in detail the region of negative $s_\alpha$, in Fig.(\ref{RATIOgammasn}) we show $\mathcal{R}_{VV}^{\textrm{SSM}}$ ($V=W$ on the left, $V=Z$ on the right) as a function of 
the singlet mass $m_S$ for three fixed values of $s_\alpha$, namely $s_\alpha=-0.09, -0.17$ (which are the two extremes of the considered range) and its central value $s_\alpha=-0.13$.
The dependence on the mixing starts to be significant  for $m_S \gtrsim 400$ GeV while it can be neglected for smaller masses, for both cases $V=W,Z$. In addition, $\mathcal{R}_{VV}^{\textrm{SSM}}$ becomes negative when the scalar mass is roughly larger than $800$ GeV, as it was the case for $s_\alpha > 0$, see Fig.(\ref{SMvsSSM}).
As before, in Fig.(\ref{gamman2}) we summarize our results for the decay widths $\Gamma(S \to Z Z )$ and $\Gamma(S \to W^+W^-(\gamma))$ as a function of $m_S$ for the selected values 
$s_\alpha=-0.17$ and $w=4.34 \,v$. Also in this case the NLO results (solid line) are very similar to the LO lines (dashed line) and tend to be larger  
in the region of small masses. 
%The effect of the gauge dependence is very similar to the case of $s_\alpha >0$. 
%even if the ratio $\Delta \Gamma_V$ is slightly reduced in the high mass region (see right plot of Fig.(\ref{Zeri})).
\section{Conclusions}
\label{concl}
In this paper we have studied in details an extension of the SM which involves the presence of a new real scalar field $s^0$, singlet under the 
SM gauge group. Its main effect is to mix with the SM scalar doublet $\phi$ via a quartic interaction of 
the form $\kappa\,(\phi^\dagger\phi) (s^0)^2$, giving rise to two mass eigenstates that we call $H$ (the lighter) and $S$ (the heaviest). 
We have limited our interests here to the study of the decay rates of $S$ to a pair of vector gauge bosons $V$; as 
far as we know, the amplitudes of such vertices could be extracted by the one-loop self energies and vertex corrections in Ref.\cite{kanemura1} (relevant to the Higgs scalar decays to gauge bosons)
but the one-loop corrections to heavy scalar decays into gauge bosons have not been computed explicitly before.
In the mass range analysed in this paper, 
$200 \leq m_S \leq 1000$ GeV which corresponds to mixing angles in the range $|s_\alpha| \in [0.09,0.17]$ and to singlet vev values $w\geq 4.34 v$, the decay $S \to VV$ is kinematically accessible and we estimated that the one-loop corrections to 
$ZZ$ and $W^+W^-(\gamma)$ channels can be as large as ${\cal O}(5\%)$ and ${\cal O}(7\%)$ for $m_S \sim 200$ GeV, respectively.
Interestingly enough, the sign of the NLO corrections  is not fixed a priori: 
for $m_S \lesssim 700$ GeV, the quantity $\mathcal{R}_{VV}^{\textrm{SSM}}=[(\Gamma^{\textrm{NLO}}_{VV}/\Gamma^{\textrm{LO}}_{VV}) -1]$ is positive for every values of $\alpha$ while for larger masses $\mathcal{R}_{VV}^{\textrm{SSM}}$ 
can also become negative (the precise turning point depends on the assumed values of $\alpha$ and $w$). Regarding the dependence on $\alpha$, $\mathcal{R}_{VV}^{\textrm{SSM}}$ exhibits a different behavior with respect 
to the sign of the mixing angle: for $\sin \alpha>0$ and fixed $m_S$, it is almost independent on $\alpha$ in the low mass range while a stronger dependence is visible in the high mass range especially for $\sin \alpha<0$.
This dependence is confined in the high mass region for masses somehow larger than $400$ GeV.
We have also studied the dependence of $\mathcal{R}_{VV}^{\textrm{SSM}}$ on the singlet vev $w$; we found that its maximum variation is practically 
the same for every $m_S$ value; to give an example, when $m_S = 900$ GeV and $s_\alpha \sim 0.2$, 
the absolute difference $|\mathcal{R}_{VV}^{\textrm{SSM}}(6.67v) - \mathcal{R}_{VV}^{\textrm{SSM}}(4.34v)|\%$ amounts to $\sim 0.3\%$ for both ratios.\\
% The corrections to the scalar singlet decay widths 
% could become relevant in EW processes with leptons in the final states, like $pp (\bar{q}q) \to S Z \to XY\,\bar{l}l$, in the hadron colliders (see Fig.(\ref{XYll})).
% \begin{figure}[h!]
% \centering
% \includegraphics[scale=0.575]{ffSZXYll}
% %\includegraphics[scale=0.475]{gammap}
% \caption{\it Feynman diagram related to the process $\bar{q}q \to S Z \to XY\,\bar{l}l$.}
% \label{XYll}
% \end{figure}
\section{Acknowledgements}
We are strongly indebted with Giuseppe Degrassi, Roberto Franceschini and Ramona Grober for illuminating discussions and Andrea Di Iura for useful suggestions on technical aspects 
of the software used in our computations.

\newpage

\appendix

\section{Feynman Rules of the Scalar Sector}
\label{SCALCOU}
We give the Feynman rules of the trilinear and quartic vertices arising from scalar potential $V_{\textrm{np}}(\phi , s^0)$, shown in eq.(2). When we expand $V_{\textrm{np}}(\phi , s^0)$ in terms of
the physical fields, the scalar trilinear and quartic couplings are generally expressed as
\begin{align}
V_{\textrm{np}}(\phi , s^0) \ = \ ... + \textrm{C}_{\mathcal{S}_1\mathcal{S}_2\mathcal{S}_3} \, \mathcal{S}_1\mathcal{S}_2\mathcal{S}_3 + 
\textrm{C}_{\mathcal{S}_1\mathcal{S}_2\mathcal{S}_3\mathcal{S}_4} \, \mathcal{S}_1\mathcal{S}_2\mathcal{S}_3\mathcal{S}_4 \,,
\end{align}
where $\mathcal{S}$ can be $H,S,\eta^3,\eta^\pm$ and the coefficients $\textrm{C}_{\mathcal{S}_1\mathcal{S}_2\mathcal{S}_3},\textrm{C}_{\mathcal{S}_1\mathcal{S}_2\mathcal{S}_3\mathcal{S}_4}$ are given by:
\begin{align}
\textrm{C}_{HHH} \ &= \ -3i c_\alpha s_\alpha (s_\alpha v - c_\alpha w)\kappa - 6i(c_{\alpha}^3 v\lambda - s_{\alpha}^3 w\rho) \,, \nonumber \\
\textrm{C}_{SSS} \ &= \ -3i c_\alpha s_\alpha (c_\alpha v + s_\alpha w)\kappa - 6i(s_{\alpha}^3 v\lambda + c_{\alpha}^3 w\rho) \,, \nonumber \\
\textrm{C}_{HSS} \ &= \ -i[c_\alpha v (c_\alpha^2-2s_\alpha^2) + s_\alpha w (2c_\alpha^2-s_\alpha^2)]\kappa - 6 i (c_\alpha s_\alpha^2 v \lambda - c_\alpha^2 s_\alpha w \rho) \,, \nonumber \\
\textrm{C}_{HHS} \ &= \ -i[s_\alpha v (s_\alpha^2-2c_\alpha^2) + c_\alpha w (c_\alpha^2-2s_\alpha^2)]\kappa - 6 i (c_\alpha^2 s_\alpha v \lambda - c_\alpha s_\alpha^2 w \rho) \,, \nonumber \\
\textrm{C}_{H\eta^3\eta^3} \ &= \  -i(2c_\alpha v \lambda -s_\alpha w \kappa) \,, \nonumber \\
\textrm{C}_{H\eta^+\eta^-} \ &= \  -i(2c_\alpha v \lambda -s_\alpha w \kappa) \,, \nonumber \\
\textrm{C}_{S\eta^3\eta^3} \ &= \  -i(c_\alpha w \kappa + 2s_\alpha v \lambda) \,, \nonumber \\
\textrm{C}_{S\eta^+\eta^-} \ &= \  -i(c_\alpha w \kappa + 2s_\alpha v \lambda) \,, \nonumber \\
\textrm{C}_{HHHH} \ &= \ -6i(c_\alpha^4 \lambda + c_\alpha^2 s_\alpha^2 \kappa + s_\alpha^4 \rho) \,, \nonumber \\
\textrm{C}_{SSSS} \ &= \ -6i(c_\alpha^4 \rho + c_\alpha^2 s_\alpha^2 \kappa + s_\alpha^4 \lambda) \,, \nonumber \\
\textrm{C}_{HHSS} \ &= \ -i(c_\alpha^4 - 4c_\alpha^2s_\alpha^2 + s_\alpha^4)\kappa - 6i c_\alpha^2s_\alpha^2(\lambda+\rho) \,, \nonumber \\
\textrm{C}_{HHHS} \ &= \ 3i c_\alpha s_{\alpha} c_{2\alpha} \kappa -6i(c_\alpha^3 s_\alpha \lambda - c_\alpha s_\alpha^3 \rho) \,, \nonumber \\
\textrm{C}_{HSSS} \ &= \ -3i c_\alpha s_{\alpha} c_{2\alpha} \kappa  -6i(c_\alpha s_\alpha^3 \lambda - c_\alpha^3 s_\alpha \rho) \,, \nonumber \\
\textrm{C}_{HH\eta^3\eta^3} \ &= \ -i(s_\alpha^2 \kappa + 2c_\alpha^2 \lambda) \,, \nonumber \\
\textrm{C}_{SS\eta^3\eta^3} \ &= \ -i(c_\alpha^2 \kappa + 2s_\alpha^2 \lambda) \,, \nonumber \\
\textrm{C}_{HS\eta^3\eta^3} \ &= \ -i s_\alpha c_\alpha (2\lambda - \kappa) \,, \nonumber \\
\textrm{C}_{HH\eta^+\eta^-} \ &= \ -i(s_\alpha^2 \kappa + 2c_\alpha^2 \lambda) \,, \nonumber \\
\textrm{C}_{SS\eta^+\eta^-} \ &= \ -i(c_\alpha^2 \kappa + 2s_\alpha^2 \lambda) \,, \nonumber \\
\textrm{C}_{HS\eta^+\eta^-} \ &= \ -i s_\alpha c_\alpha (2\lambda - \kappa) \,. \nonumber
\end{align}
The couplings of $H$ and $S$ with the Standard Model ordinary matter are:
\begin{align}
\textrm{C}_{H\sigma \bar{\sigma}} \ &= \  c_\alpha \, \textrm{C}_{h\sigma \bar{\sigma}} \,, \ \ \ \textrm{C}_{HH\sigma \bar{\sigma}} \ = \ c_\alpha^2 \, \textrm{C}_{hh\sigma \bar{\sigma}} \,, \nonumber \\
\textrm{C}_{S\sigma \bar{\sigma}} \ &= \  s_\alpha \, \textrm{C}_{h\sigma \bar{\sigma}} \,, \ \ \ \textrm{C}_{SS\sigma \bar{\sigma}} \ = \ s_\alpha^2 \, \textrm{C}_{hh\sigma \bar{\sigma}} \,, \nonumber
\end{align}
where $\textrm{C}_{h \sigma \bar{\sigma}}, \textrm{C}_{hh\sigma \bar{\sigma}}$ are the couplings of the Standard Model Higgs field ($h$) with a pair of bosonic or fermionic fields ($\sigma=W,Z,f$).

\newpage

\section{Feynman Diagrams for $S \to VV$}
\label{FeynDiag}
\subsection*{Contribution to the Self Energies}
\begin{figure}[h!]
\centering
\includegraphics[scale=0.435]{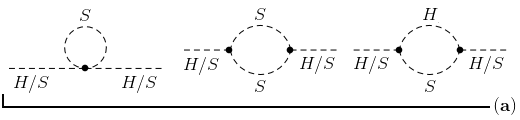}
\includegraphics[scale=0.435]{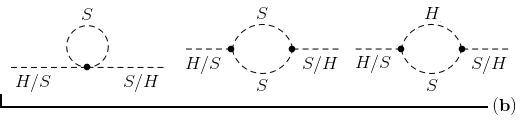}
\caption{\it  ${\rm Fig.(a)}$: S field contributions to the scalar self-energies; ${\rm Fig.(b)}$: S field contributions to the mixed scalar self-energies.}
\includegraphics[scale=0.435]{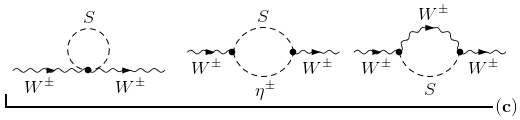}
\includegraphics[scale=0.435]{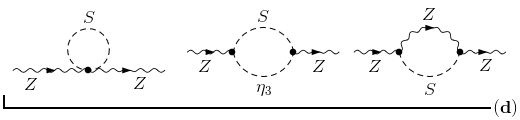}
\caption{\it ${\rm Fig.(c)}$: S field contributions to the $W$ boson self energies; ${\rm Fig.(d)}$: S field contributions to the $Z$ boson self energies.}
\end{figure}
\subsection*{Contribution to the Trilinear Vertices}
\begin{figure}[h!]
\centering
\includegraphics[scale=0.45]{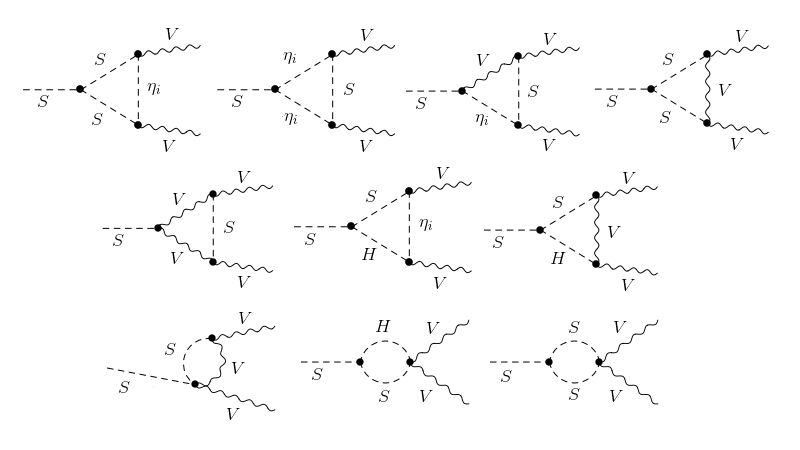}
\caption{\it S field contribution to the $SVV$ vertex.}
\end{figure}

\newpage

\section{Bremsstrahlung}
\label{BREM}
Here we explicitly report the photon bremsstrahlung contributions to the process $S \to W^+W^-(\gamma)$. As shown in eq.(\ref{gbrem}), the total bremsstrahlung decay rate is given by the sum of soft- and hard-photon factors which can be written as,
\begin{align}
\label{dsoft}
\delta^{\textrm{soft}}_{W} \ &= \ \frac{\alpha_{em}}{\pi}\bigg\{ \mathcal{G}(r) \ln\left(\frac{4\Lambda_\gamma^2}{m_\gamma^2}\right) + \left(\mathcal{G}(r)+1 \right)\left[\frac{\mathcal{N}_1}{\textrm{a}_1} + \frac{2}{2r-1}\right]\bigg\} \,, \\
\label{dhard}
\delta^{\textrm{hard}}_{W} \ &= \ \frac{\alpha_{em}}{\pi}\bigg\{ \mathcal{G}(r) \ln\left(\frac{m_S^2}{4\Lambda_\gamma^2}\right) + \frac{\mathcal{N}_2\left(\mathcal{G}(r)+1 \right)}{\textrm{a}_1}
+ \frac{14}{3}\left(1 - \frac{t}{r}\sqrt{1-\frac{1}{t}} \, \right) \, + \nonumber \\ 
& \ \ \ +\frac{1}{\mathcal{N}_3}\left[\frac{\textrm{a}_2-\textrm{a}_1}{2r^2} \left(2 -\frac{1}{r}\right) + \frac{t}{3r}\sqrt{1-\frac{1}{t}} \left(1- \frac{t}{r} \right)\left(2-\frac{4t-1}{r} \right)\right] \, - \nonumber \\
& \ \ \ \ \ \ \ \ \ \ \ \ \ \ \ \ \ \ \ \ \ \ \ \ \ \ \ \ \ \ \ \ \ \ - 2\ln\left( \frac{1-(\textrm{c}_-\textrm{d}_+)^2}{1-(\textrm{c}_-\textrm{d}_-)^2}\right) + 4\left(\frac{t}{r}\textrm{a}_2 - \textrm{b}_- \right) \bigg\} \,,
\end{align}
where,
\begin{align}
\mathcal{G}(r) \ &= \ [\, \textrm{a}_1\,(2-1/r)/(\sqrt{1-1/ r}) \,]-1 \,, \nonumber \\
\mathcal{N}_1  \ &= \ \textrm{Li}_2(\textrm{c}_-^2)+\textrm{a}_1\,(\textrm{a}_1-2\textrm{b}_-) -(\pi^2/6) \,, \nonumber \\
\mathcal{N}_2  \ &= \ \textrm{Li}_2((\textrm{c}_-\textrm{d}_+)^2) - \textrm{Li}_2((\textrm{c}_-\textrm{d}_-)^2) + \textrm{Li}_2(\textrm{c}_-)^4 + 4\textrm{a}_1\,(\textrm{a}_1-\textrm{b}_+) \, + \nonumber \\
& \ \ \ \ \ \ \ \ \ \ \ \ \ \ \ \ \ \ \ +2\textrm{a}_2\ln[(1-(\textrm{c}_-\textrm{d}_+)^2)(1-(\textrm{c}_-\textrm{d}_-)^2)]  -(\pi^2/6) \,, \nonumber \\
\mathcal{N}_3  \ &= \ \sqrt{1-1/r}\left[\, 1 - (1/r) + (3/4r^2) \, \right] \,, \nonumber \\
\textrm{a}_1   \ &= \ \ln(\textrm{c}_+) \,, \nonumber \\ 
\textrm{a}_2   \ &= \ \ln(\textrm{d}_+) \,, \nonumber \\
\textrm{b}_\pm \ &= \ \ln(\textrm{c}_+\pm\textrm{c}_-) \,, \nonumber \\
\textrm{c}_\pm \ &= \ \sqrt{r}\pm\sqrt{r-1} \,, \nonumber \\ 
\textrm{d}_\pm \ &= \ \sqrt{t}\pm\sqrt{t-1} \,, \nonumber
\end{align}
with $r = m_S^2/4m_W^2$ and $t = r(1-2q_\gamma/m_S)$. For $q_\gamma=q_\gamma^{\textrm{max}}$, eq.(\ref{dhard}) is reduced to,
\begin{align}
\delta^{\textrm{hard}}_{W} \ &= \ \frac{\alpha_{em}}{\pi}\bigg\{ \mathcal{G}(r) \ln\left(\frac{m_S^2}{4\Lambda_\gamma^2}\right) - 4\textrm{b}_- + \frac{14}{3} \, +\nonumber \\ 
& \ \ \ + \frac{\left(\mathcal{G}(r)+1 \right)}{\textrm{a}_1}\left[\textrm{Li}_2(\textrm{c}_-)^4 -\frac{\pi^2}{6} + 4\textrm{a}_1\,(\textrm{a}_1-\textrm{b}_+) - \frac{2\textrm{a}_1}{4r^2-4r+3}\right]\bigg\}. 
\end{align}

\newpage

\bibliographystyle{MLA}

\end{document}